\input epsf.tex
\documentclass[12pt]{article}
\begin{document}
\begin {titlepage}
\begin{flushright} ULB-TH/02-09\\   hep-th/0203097\\
\end{flushright}
\vspace{1cm}
\begin{center} {\large \bf  A Brief Course in Spontaneous Symmetry 
Breaking \\ II.  Modern Times: The BEH Mechanism}\footnote{Invited talks
presented at the 2001 Corfu Summer Institute on Elementary Particle Physics}\\
\vspace{1cm}  Fran\c cois Englert\footnote{ E-mail : fenglert@ulb.ac.be}
\\
\vspace{.5cm} {\it Service de Physique Th\'eorique}\\ {\it Universit\'e
Libre de Bruxelles, Campus Plaine, C.P.225}\\ {\it Boulevard du Triomphe,
B-1050 Bruxelles, Belgium}\\
\end{center}
\vspace{3cm}
\begin{abstract}

\noindent The theory of symmetry breaking in  presence of gauge   fields
is presented, following the historical track.  Particular emphasis is
placed upon the underlying concepts.

\end{abstract}
\end {titlepage}
\addtocounter{footnote}{-2}

\parskip 10pt plus 1pt
\setlength{\parindent}{0cm} {\bf \large I. Introduction}
\medskip

It was known in the first half of the twentieth century that, at  the atomic
level and at larger distance scales, all phenomena appear to be governed by
 the  laws of classical general     relativity and  of quantum
electrodynamics. 

Gravitational and electromagnetic forces are long range and  hence can be
perceived directly without the mediation of highly sophisticated
technical devices.   The development of large scale physics, initiated by 
the Galilean inertial principle, is surely tributary to this circumstance.
It   then took about three centuries to achieve a  successful description of
long range effects.

The discovery of subatomic structures and of the concomitant weak  and
strong interaction short range forces   raised the question of how to cope
with  short range forces in quantum field theory. The Fermi theory of
weak   interactions, formulated in terms of a four Fermi point-like
current-current interaction, was predictive in lowest order perturbation
theory and successfully confronted many experimental data. However, it
was clearly inconsistent in higher order because of uncontrollable
quantum divergences at high energies. In order words, in contradistinction
with quantum electrodynamics, the Fermi theory is not renormalizable.
This difficulty could not be solved by smoothing the point-like interaction
by a massive, and therefore short range, charged vector particle exchange
(the so-called $W^+$ and $W^-$ mesons); theories with fundamental
massive  charged vector mesons are  not renormalizable either. In the
early nineteen sixties, there seemed to be insuperable obstacles for 
formulating  a  theory with short range forces mediated by massive
vectors.

The solution of the latter problem came from the theory proposed in 1964
by Brout and Englert \cite{eb} and by Higgs \cite{higgs1,higgs2}. The
Brout-Englert-Higgs (BEH) theory is based on a mechanism, inspired from
the spontaneous  symmetry breaking of a continuous symmetry, discussed
in the previous talk by Robert Brout,    adapted to gauge  theories and in
particular to non abelian gauge theories. The mechanism unifies   long
range and short range forces mediated by vector mesons,  by deriving the
vector mesons masses from a fundamental theory containing only
massless vector fields. It led to a solution of the weak interaction puzzle
and opened the way to  modern perspectives on unified  laws of nature.

Before turning to an expos\'e of the BEH mechanism, we shall in section II
review,  in the context of quantum field theory,  the analysis given by
Robert Brout of the spontaneous breaking of a continuous symmetry.
Section III  explains the BEH mechanism. We  present the quantum field
theory approach of Brout and Englert wherein the breaking mechanism for
both abelian and non abelian gauge groups is induced  by scalar  bosons. We
also present their approach in the case of  dynamical symmetry breaking 
from  fermion condensate. We then turn to the equation of motion
approach of Higgs. Finally we explain the renormalization issue. In section
IV, we briefly review the well-known applications of the BEH mechanism
with particular emphasis on concepts relevant to the quest for unification.
Some comments on this subject are made in section V.
\bigskip\bigskip

 {\bf \large II. Spontaneous Breaking of a Global Symmetry}
\medskip

Spontaneous   breaking of a Lie group symmetry was discussed by Robert
Brout in ``The Paleolitic Age''.  I  review here its essential features  in the
quantum field theory context.

Recall that spontaneous breakdown of a continuous  symmetry in
condensed matter physics implies a degeneracy of the ground state, and
as  a consequence, in absence of long range forces, collective modes
appear whose energies go to zero when the wavelength goes to infinity. This
was  exemplified in particular by  spin waves in a Heisenberg
ferromagnet. There, the broken symmetry  is the rotation invariance.  

Spontaneous symmetry breaking was introduced in relativistic quantum
field theory by Nambu  in analogy to the BCS theory of
superconductivity.   The problem studied by Nambu
\cite{nambu1} and Nambu and Jona-Lasinio \cite{nambujl} is the
spontaneous breaking of chiral symmetry induced by a fermion 
condensate\footnote{See the detailed discussion in Brout's
lecture, section VII.}. The chiral phase group
$\exp (i\gamma_5 \alpha)$  is broken by the fermion condensate $\langle
\bar\psi\psi\rangle \neq 0$ and the massless mode is identified with the
pion. The latter  gets its tiny mass (on the hadron scale) from a small
explicit breaking of the symmetry, just as a small external magnetic field
imparts a small gap in the spin wave spectrum. This interpretation of the
pion mass constituted a breakthrough in our understanding of strong
interaction physics. General features of spontaneous symmetry breakdown
in relativistic quantum field theory were further formalized by Goldstone
\cite{goldstone}. Here, symmetry is broken by non vanishing  vacuum
expectation values of scalar fields.   The method is designed to exhibit
the appearance of a  massless  mode out of the degenerate  vacuum and
does not really depend on the significance of the scalar fields. The latter
could be elementary or represent  collective  variables of   more
fundamental fields, as would be the case in the original Nambu model.
Compositeness  affects details of the model considered, such as the
behavior at high momentum transfer,  but  not  the existence of the
massless excitations encoded in the degeneracy of the vacuum.
 
Let us first illustrate the occurrence of this massless Nambu-Goldstone
(NG) boson in a simple model of a complex scalar field with $U(1)$
symmetry \cite{goldstone}.

The Lagrangian density,
\begin{equation}
\label{global} {\cal L} =\partial ^\mu\phi^*\partial_\mu\phi
-V(\phi^*\phi)
\quad\hbox{with} \quad V(\phi^*\phi)= -\mu^2 \phi^*\phi  + \lambda
(\phi^*\phi)^2~,~\lambda > 0\ ,
\end{equation}  is invariant under the  $U(1)$ group $\phi\to
\displaystyle {e^{i\alpha} \phi}$.  The $U(1)$ symmetry is called global
because the group parameter $\alpha$ is  constant in space-time. It is
broken by a vacuum expectation value of the
$\phi$-field given, at the classical level, by the minimum of
$V(\phi^*\phi)$. Writing
$\phi = (\phi_1 + i
\phi_2)/ \sqrt2$, one may choose $\langle
\phi_2\rangle=0$. Hence $ \langle
\phi_1\rangle^2=\mu^2/\lambda $ and we select, say,  the vacuum with 
$\langle
\phi_1\rangle
 $  positive.  The potential $V(\phi^*\phi)$ is depicted in Fig.1 .  

\hskip 2.5cm\epsfbox{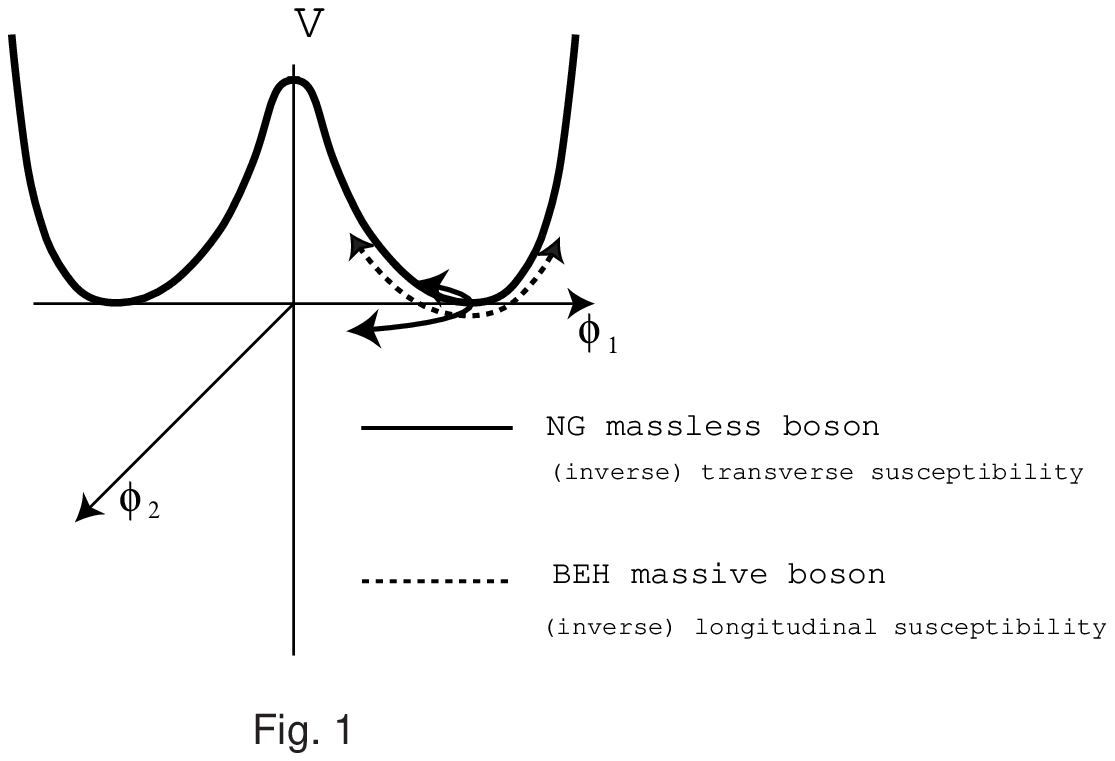}

Around the unbroken vacuum the field $\phi_1$ has negative mass and
acquires a positive mass around the broken vacuum where the field
$\phi_2$ is massless. The latter is the NG boson of broken U(1) symmetry.
The massive scalar describes the fluctuations of the order parameter
$\langle
\phi_1\rangle$. Its mass is the analog of the inverse longitudinal
susceptibility of the  Heisenberg ferromagnet discussed by Robert Brout
while the vanishing of the NG boson mass corresponds to the vanishing of
its   inverse  transverse susceptibility.  The scalar boson
$\phi_1$ is always present in spontaneous breakdown of a   symmetry. In
the context of the BEH mechanism analyzed in the following section, it
was introduced by Brout and myself, and by Higgs. We shall label it the BEH
boson\footnote{It is
  often called the Higgs boson in the literature.} (Fig.1).

In the classical limit, the origin of the massless NG boson $\phi_2$ is
clearly illustrated in the Fig.1. The vacuum characterized by the order
parameter $\langle\phi_1\rangle$ is rotated into an equivalent vacuum by
the field  $\phi_2$ at zero space momentum. Such rotation costs no
energy and thus the field $\phi_2$ at space momenta $\stackrel{\to}{q}
=0$ has $q_0=0$  on the equations of motion, and   hence  zero mass.

 This can be formalized and generalized by noting that the conserved
Noether current
 $J_\mu= \phi_1\partial_\mu \phi_2 -\phi_2\partial_\mu \phi_1$ gives a
charge $ Q =\int  J_0 d^3x$. The operator $\exp\, (i\alpha Q)$ rotates the
vacuum by an angle $\alpha$. In the classical limit, this charge is, around
the chosen vacuum, $ Q =\int  \langle\phi_1\rangle \partial_0 \phi_2
d^3x$ and involves only $\phi_2$ at zero momentum. In general, 
$\langle[Q, \phi_2]\rangle = i
\langle\phi_1\rangle $ is non zero in the chosen vacuum. This implies that
the propagator $\partial^\mu  \langle T J_\mu(x) ~\phi_2
(x^\prime)\rangle$ cannot vanish at zero four-momentum $q$ because its
integral over space-time is precisely $\langle[Q, \phi_2]\rangle $.
Expressing the propagator in terms of Feynman diagrams we see that  the
$\phi_2$-propagator must have a pole at $q^2=0$. The field $\phi_2 
$ is the massless NG boson.

The proof is immediately extended to the  spontaneous  breaking of a
semi-simple Lie group global symmetry. Let $\phi^ A$ be scalar fields
spanning a representation of the Lie group
$\cal G$ generated by the (antihermitian) matrices
$T^{aAB}$. If the dynamics is governed by a $\cal G$-invariant action and
if the potential has  minima for non vanishing $\phi^ A$,s ,  symmetry is
broken  and the vacuum is degenerate under $\cal G
$-rotations. The conserved charges are $ Q^a=\int 
\partial_\mu\phi^ B~ T^{aBA}~\phi^ A ~d^3x$. As in the abelian case above,
the propagators of the fields $\phi^B$ such that $\langle[Q^a,
\phi^B]\rangle = T^{aBA}~\langle\phi^ A\rangle \neq 0$   have a NG pole at
$q^2=0$.
\bigskip\bigskip

{\bf \large III. The BEH Mechanism}
\bigskip

{\bf - From global to local symmetry}

The global $U(1)$ symmetry in Eq.(\ref{global})   can be extended  to a
local $U(1)$ invariance $\phi(x)\to
\displaystyle {e^{i\alpha(x)} \phi(x)}$ by introducing a vector field
$A_\mu(x)$ transforming according to  $A_\mu(x)\to A_\mu(x) + (1/ e)
\partial_\mu \alpha(x)$. The  corresponding Lagrangian density is
\begin{equation}
\label{local} {\cal L} =D^\mu\phi^*D_\mu\phi -V(\phi^*\phi) -{1\over4}
F_{\mu\nu} F^{\mu\nu} \ ,
\end{equation} with covariant derivative $D_\mu\phi =
\partial_\mu\phi -ieA_\mu
\phi $ and $F_{\mu\nu} = \partial_\mu A_\nu -\partial_\nu A_\mu $.

Local invariance under a semi-simple Lie group $\cal G$ can be realized by
extending  the Lagrangian  Eq.(\ref{local}) to incorporate  non-abelian
 Yang-Mills vector fields $A_\mu^a$ 
\begin{eqnarray}
\label{localym} &&{\cal L}_{\cal G} =  ( D ^\mu\phi)^{*A} (D_\mu\phi)^A
-V  -\displaystyle { {1\over 4}} F_{\mu\nu}^aF^{a\,\mu\nu}\ ,\\
&&\hbox{where}\nonumber\\
\label{covariant} &&(D_\mu \phi)^A =\partial_\mu \phi^A -  eA_\mu^a
T^{a\,AB }\phi^B , F_{\mu\nu}^a =\partial_\mu A_\nu^a -\partial_\nu
A_\mu^a -e f^{abc}A_\mu^b A_\nu^c\ . 
\end{eqnarray} Here, $\phi^A$ belongs to the representation of
$\cal G$ generated by
$T^{a\,AB }$ and the potential $V$ is invariant under  $\cal G$.

The success of quantum electrodynamics based on local U(1) symmetry,
and of classical general relativity based on a local generalization of
Poincare invariance, provides ample evidence for the relevance of local
symmetry for the description of natural laws.  One expects that local
symmetry has a fundamental significance rooted in causality and in the
existence of exact conservation laws at a fundamental level, of which 
charge conservation appears as the prototype. As an example of the
strength of local symmetry we cite the fact that conservation laws
resulting from a global symmetry alone are violated in presence of black
holes.  

The local symmetry, or gauge invariance, of Yang-Mills theory, abelian or
non abelian, {\em apparently} relies on the massless character of the
gauge  fields $A_\mu$, hence on the long range character of the forces
they transmit,  as the addition of a mass term for $A_\mu$ in the
Lagrangian Eq.(\ref{local}) or (\ref{localym})  destroys gauge invariance.
But   short range forces, such as the weak interaction forces, 
  seem to be as fundamental as the electromagnetic ones despite the
apparent absence  of exact conservation laws. To reach a basic
description of such forces one is  tempted to link the violation of
conservation to a mass of the gauge fields which would arise from
spontaneous symmetry breaking.  However the problem of spontaneous
broken  symmetry is  different for global and for local symmetry.

To understand the difference, let us break the symmetries explicitly. To
the Lagrangian Eq.(\ref {global}) we add the term
\begin{equation}
\label{breakglobal}
\phi h^* +\phi^* h\ ,
\end{equation} where $h , h^*$ are constant in space time. Let us take h
real. The presence of the  field $h$ breaks explicitly the global $U(1)$
symmetry and the field
$\phi_1$ always develops an expectation value. When $h\to 0$, the
symmetry of the action is restored but, when the symmetry is broken by a
minimum of $V(\phi\phi^*)$ at $\vert\phi\vert   \neq 0$, we still have
$\langle
\phi_1\rangle \neq 0$. The tiny
$h$-field  simply picks up one of the degenerate vacua  in perfect analogy
with the infinitesimal magnetic field which orients the magnetization of
a ferromagnet. As in statistical mechanics, spontaneous broken global
symmetry can be recovered in the limit of vanishing external symmetry
breaking. The degeneracy of the vacuum can be put into evidence by
changing the phase of $h$; in this way, we can  reach in the limit
$h\to 0$ any  $U(1)$ rotated vacuum.

When the symmetry is extended from global to local, one can still break
the symmetry by an external ``magnetic'' field. However in the limit of
vanishing magnetic field the expectation value of any gauge dependent
local operator will tend to zero  because, in contradistinction to global
symmetry, it  cost no energy in the limit to change the relative orientation
of neighboring ``spins''; there is then no ordered configuration in group
space which can be protected from disordering fluctuations. As a
consequence, the vacuum is generically non degenerate and points in no
particular direction in group space as the external field goes to zero. Local
gauge symmetry cannot be spontaneously broken\footnote{For a detailed
 proof, see reference
\cite{elitzur}.} and the vacuum is gauge invariant\footnote{Note that for
global symmetry breaking, one can always choose a linear combination of
degenerate vacua which is invariant under, say, the
$U(1)$ symmetry. This  choice  has no observable consequences and  only
masks the degeneracy of the vacuum which is guaranteed by a
superselection rule. The Hilbert space splits indeed, as in the
ferromagnetic case analyzed by Robert Brout (section V of ``The
Paleolitic Age''), into an infinite number of orthogonal spaces formed by all
the finite excitations on each  degenerate vacuum.}.  Recalling that the
explicit presence of a gauge vector mass
 breaks gauge invariance, we are thus faced with a dilemma.  How can 
gauge fields acquire mass
 without breaking the local symmetry?
\bigskip

 {\bf - Solving the dilemma}

In perturbation theory, gauge invariant quantities are evaluated by
choosing a particular gauge.  One  imposes the gauge condition by adding
to the action a gauge fixing term and  one sums over subsets of graphs
satisfying the Ward Identities\footnote{To this end, it is  often
necessary, in particular for non abelian gauge theories, to include
Fadeev-Popov ghosts terms in the action. These  contribute  when closed
gauge  field loops are included in the computation.}.

Consider the Yang-Mills theory defined by the Lagrangian
Eq.(\ref{localym}). Let us  choose a gauge which preserves Lorentz
invariance and a residual global $\cal G$ symmetry. This can be achieved
by adding to the Lagrangian a gauge fixing term $(2\eta)^{-1} \partial_\mu
A_a^\mu \, \partial_\nu A^{a\, \nu} $. The gauge parameter $\eta$ is 
arbitrary and has no observable consequences.

\vskip 1cm 
\hskip .7cm
\epsfbox{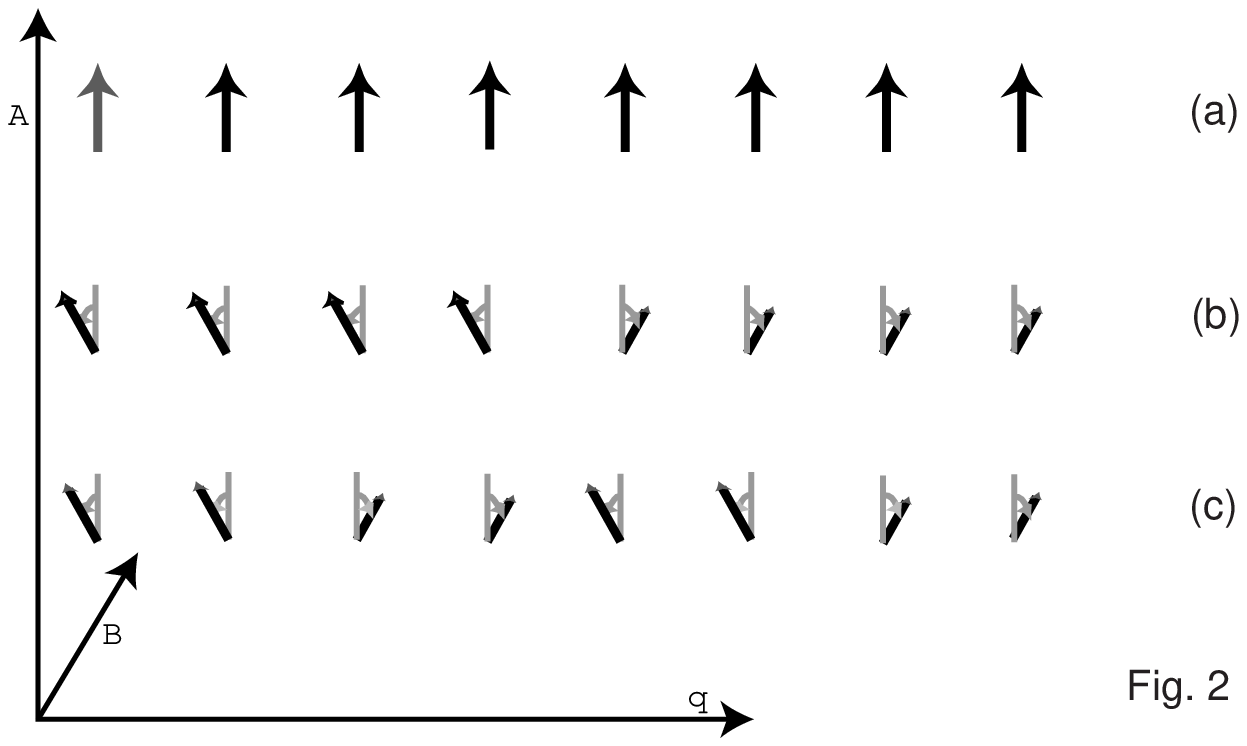}

The global symmetry can now be spontaneously broken, for suitable
potential~$V$, by  non zero expectation values $\langle
\phi^A\rangle$ of BEH fields. In Fig.2 we have represented   fluctuations of
this parameter in the spatial $q$-direction and  in  an internal space
direction  orthogonal to the    direction $A$. The  orthogonal direction
depicted in the figure has been labeled  $B$. Fig.2a pictures the
spontaneously broken vacuum of the  gauge fixed Lagrangian. Fig.2b and 2c
represent fluctuations of finite wavelength
$\lambda$.

Clearly as
$\lambda\to \infty$ these fluctuations can only induce global rotations in
the internal space. In absence of gauge   fields, such fluctuations would
give rise, as in  spontaneously broken  global continuous symmetries, to 
massless NG mode. In a gauge theory,  fluctuations of
$\langle\phi^A\rangle$ are just   local rotations in the internal space and
hence are  unobservable gauge fluctuations.  Hence the NG bosons induce
only gauge transformations and its  excitations disappear from the
physical  spectrum.

The degrees of freedom  of the NG fields  were present in the original
gauge invariant action and cannot disappear.   But what makes local
internal space rotations    unobservable in a gauge theory is precisely the
fact that they can be absorbed through  gauge transformations by the
Yang-Mills fields. The absorption of the long range NG fields
renders massive those 
gauge fields to which they are coupled, and transfers to them the missing
degrees of freedom which becomes their
third polarization. 

We shall see in the next sections how these  considerations  are realized
in quantum field  theory, giving rise to an apparent breakdown of
symmetry:   despite the absence of spontaneous local symmetry breaking,
gauge invariant vector masses will be generated in a coset
$\cal G/ H$, leaving long range forces only in a subgroup $\cal H$ of $\cal
G$.

\bigskip {\bf - The quantum field theory approach} \cite {eb}
\medskip

{\it $\alpha$) Breaking by BEH bosons}

Let us first examine the abelian case as realized by the complex scalar
field $\phi$  exemplified in Eq.(\ref {local}).

In the covariant gauges, the free propagator of the  field
$A_\mu$ is
\begin{equation}
\label {dabelian} D_{\mu\nu}^0 ={g_{\mu\nu}-q_\mu q_\nu /q^2\over q^2}
+ \eta {q_\mu q_\nu/q^2 \over q^2}\ ,
\end{equation} where $\eta$ is the  gauge parameter.  It can be put equal
to zero, as  in the Landau gauge used in reference
\cite{eb},  but we leave it  arbitrary  here to illustrate explicitly the
 role of the NG-boson.
\vskip 1cm
\hskip 2cm
\epsfbox{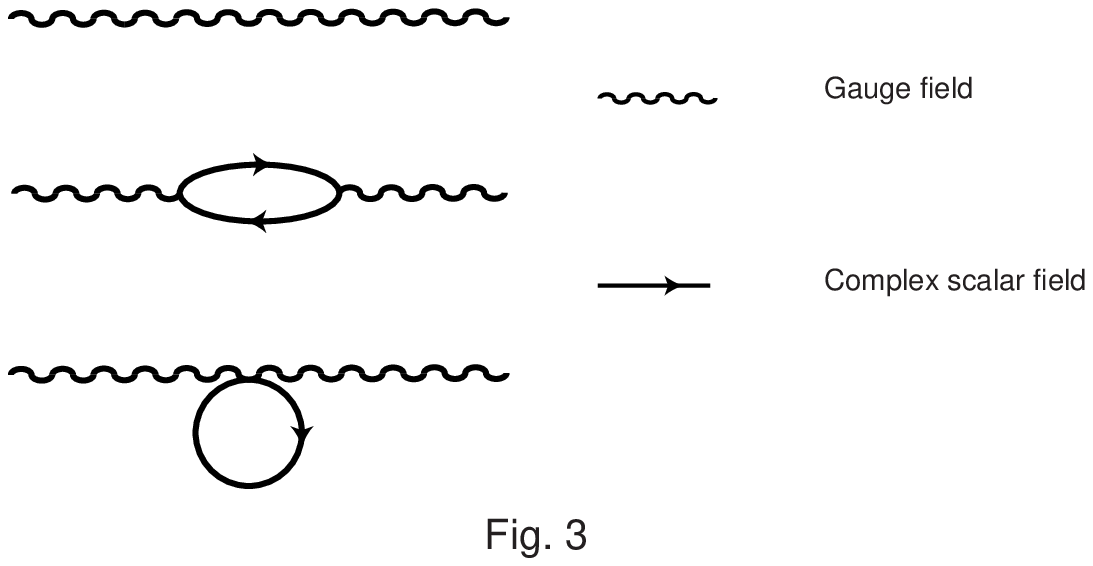} 

\vskip .5cm In absence of symmetry breaking, the lowest order
contribution to the self-energy, arising from the covariant derivative
terms in Eq.(\ref{local}), is given by the one-loop diagrams of Fig.3. The
self-energy (suitably regularized) takes the form of a polarization tensor 
\begin{equation}
\label{pabelian}
\Pi_{\mu\nu}= (g_{\mu\nu} q^2-q_\mu q_\nu)~ \Pi(q^2)\ ,
\end{equation} where the scalar polarisation $\Pi(q^2)$ is regular at
$q^2=0$, leading to the gauge  field propagator
\begin{equation} D_{\mu\nu} ={g_{\mu\nu}-q_\mu q_\nu /q^2\over q^2[1-
\Pi(q^2)]} +\eta {q_\mu q_\nu/q^2\over q^2}\ . 
\end{equation} The  polarization tensor in Eq.(\ref{pabelian}) is
transverse and hence does not affect the gauge parameter 
$\eta$.  The transversality of the polarization tensor reflects the gauge
invariance of the theory\footnote{The transversality of polarisation
tensors is a consequence of the Ward Identities alluded to in the
preceding section.} and, as we shall see below, the regularity of the
polarization scalar signals the absence of symmetry breaking. This
guarantees that the  
$A_\mu$-field remains massless.

Symmetry breaking adds tadpole diagrams to the previous ones. To see
this write
\begin{equation}
\label{vacuum}
\phi ={1\over \sqrt 2}(\phi_1 + i\phi_2)~~~
\langle
\phi_1\rangle
\neq 0\ .
\end{equation} The BEH field is $\phi_1$ and the NG field
$\phi_2$. The additional diagrams are depicted in Fig.4.
\vskip .5cm
\hskip 2cm
\epsfbox{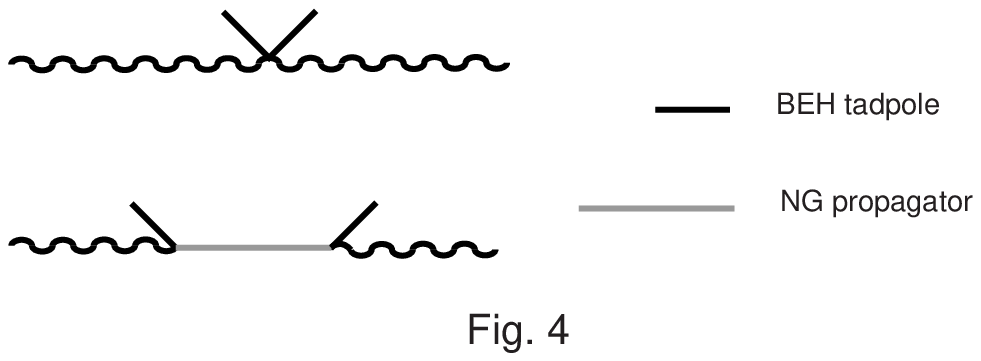}
\vskip .5cm

In this case, the polarisation scalar $\Pi(q^2)$ in Eq.(\ref{pabelian})  
acquires a pole
\begin{equation}
\label{pamass}
\Pi(q^2)={e^2\langle \phi_1\rangle^2\over q^2}\ ,
\end{equation} and, in lowest order perturbation theory,  the gauge  field
propagator becomes
\begin{equation}
\label{damass} D_{\mu\nu} ={g_{\mu\nu}-q_\mu q_\nu/q^2
\over q^2- \mu^2}  +\eta {q_\mu q_\nu/q^2\over q^2}\ , 
\end{equation} which shows that the $A_\mu$-field gets a mass
\begin{equation}
\label{massa}
\mu^2=e^2 \langle
\phi_1\rangle^2\ .
\end{equation}

The generalization of Eqs.(\ref{pabelian}) and (\ref{pamass}) to the non
abelian case described by the action Eq.(\ref{localym}) is straightforward.
One gets from the graphs depicted in Fig.5,

\hskip 3cm\epsfbox{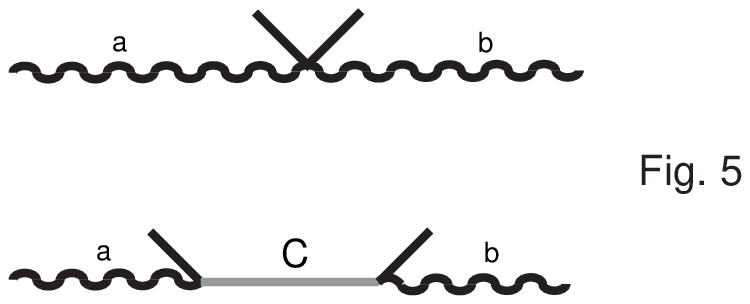}
\begin{eqnarray}
\label{pnamass} &&\Pi_{\mu\nu}^{ab}= (g_{\mu\nu}q^2-q_\mu q_\nu )
\Pi^{ab}(q^2)\ , \hskip 3cm\\
\label{pol} && 
\Pi^{ab}(q^2)={e^2\langle \phi^{*B}\rangle
T^{*a\,BC}T^{b\,CA}\langle\phi^A\rangle \over q^2}\ ,
\end{eqnarray} from which follows the mass matrix
\begin{equation}
\label{massna}
\mu^{ab} = e^2\langle \phi^{*B}\rangle
T^{*a\,BC}T^{b\,CA}\langle\phi^A\rangle\ . 
\end{equation} In terms of the non-zero eigenvalues
$\mu_a$ of the mass matrix the propagator for the massive gauge vectors
takes the same form as Eq.(\ref{damass})
\begin{equation}
\label{dnamass} D_{\mu\nu}^a  ={g_{\mu\nu}-q_\mu q_\nu/q^2
\over q^2- \mu^{a2}}  +\eta {q_\mu q_\nu/q^2\over q^2}\ . 
\end{equation}

The gauge invariance is expressed, as it was in absence of symmetry
breaking, through the transversality of the polarization tensors
Eqs.(\ref{pabelian}) and (\ref{pnamass}).  The singular  $1/q^2$
contributions to
 the polarization scalars  Eqs.(\ref{pamass}) and (\ref{pol}), which
preserve transversality while giving mass to the gauge fields, stem from
the long range  NG boson fields encoded in their $1/q^2$
propagator.  We shall
verify below that this pole has no  observable effect as such. On the other
hand,  its absorption in the gauge field propagator transfers  the degrees
of freedom of the NG bosons to the third degree of polarization  of the
massive vectors. Indeed, on the mass shell
$q^2= \mu^{a2}$, one easily verifies that the numerator in their propagator
Eq.(\ref {dnamass}) is:
\begin{equation}
g_{\mu\nu}-{q_\mu q_\nu\over q^2}=\sum_{\lambda=1}^3
e_\mu^{(\lambda)}.e_\nu^{(\lambda)}~~ , ~~ q^2=
\mu^{a2} 
\ ,
\end{equation} where the $e_\mu^{(\lambda)}$ are the three polarization
vectors which are orthonormal in the rest frame of the particle.  

In this way, the NG bosons generate massive propagators for those gauge
fields to which they are coupled. Long range forces only survive in the
subgroup
$\cal H$ of
$\cal G$ which leaves invariant the non vanishing expectation values
$\langle\phi^A\rangle$.

Note that (as in the abelian case) the scalar potential $V$  does not enter
the computation of the gauge field propagator. This is because the
trilinear term arising from the covariant derivatives  in the Lagrangian
Eq.(\ref{localym}), which yields the second graph of Fig.5,  can only couple the
tadpoles to  other scalar fields  through group rotations and hence
couple them only  to the NG bosons. These are the eigenvectors with zero
eigenvalue of the scalar mass matrix given by the quadratic term in the
expansion of the potential $V$ around its minimum. Hence the mass matrix
decouples from the tadpole at the tree level considered above.  An
explicit example of this feature will be given for the Lagrangian
Eq.(\ref{reduction}).

\medskip

{\it $\beta$) Dynamical symmetry breaking}

The symmetry breaking giving mass to gauge vector bosons may arise
from the fermion condensate breaking chiral symmetry. This is illustrated
by the following chiral invariant Lagrangian
\begin{equation} {\cal L} = {\cal L}_0^F -e_V\,\bar\psi\gamma_\mu\psi
V_\mu  -e_A\,\bar\psi\gamma_\mu\gamma_5\psi A_\mu  -  {1\over 4}
F_{\mu\nu}F^{\mu\nu}(V) - {1\over 4} F_{\mu\nu}F^{\mu\nu}(A)\ .
\end{equation} Here $F_{\mu\nu}(V)$ and $F_{\mu\nu}(A)$ are abelian
field strength for
$U(1)\times U(1)$ symmetry.  Chiral anomalies are eventually canceled by
adding in the required additional fermions.

The Ward identity for the chiral current
\begin{equation} q^\mu \Gamma_{\mu 5} (p+q/2, p-q/2) =
S^{-1}(p+q/2)\gamma_5 +\gamma_5S^{-1}(p-q/2)\ ,
\end{equation} shows that if  the fermion self-energy
$\gamma^\mu p_\mu
\Sigma_2(p^2) - \Sigma_1(p^2)
$ acquires  a non vanishing
$\Sigma_1(p^2)
$ term, thus a dynamical mass $m$ at $\Sigma_1(m^2)=m$ (taking for
simplicity
$\Sigma_2(m^2)=1), 
$ the axial vertex
$\Gamma_{\mu 5}$ develops a pole at $q^2=0$.  In leading order in
$q$, we get
\begin{equation}
\label{vertex}
\Gamma_{\mu 5}{\rightarrow} 2m\gamma_5 {q_\mu\over q^2}
\ .
\end{equation}

The pole in the vertex function induces a pole in the suitably regularized
gauge invariant polarization tensor
$\Pi_{\mu\nu}^{(A)}$ of the axial vector field $A_\mu$ depicted in Fig.6 
\begin{equation}
 \Pi_{\mu\nu}^{(A)}=  e_A^2 (g_{\mu\nu}q^2-q_\mu q_\nu )\Pi^{(A)}(q^2)\ ,
\end{equation}  with
\begin{equation}
\label{pdyn}
\lim_{q^2\to 0} q^2\Pi^{(A)}(q^2) =\mu^2
\neq 0\ .
\end{equation} The field $A_\mu$  acquires in this
approximation\footnote {The validity of the approximation, and in fact of
the dynamical approach, rests on the high momentum  behavior of the
fermion self energy, but this problem will not be discussed here.} a gauge
invariant mass
$\mu$ .

\vskip .5cm
\hskip 2cm
\epsfbox{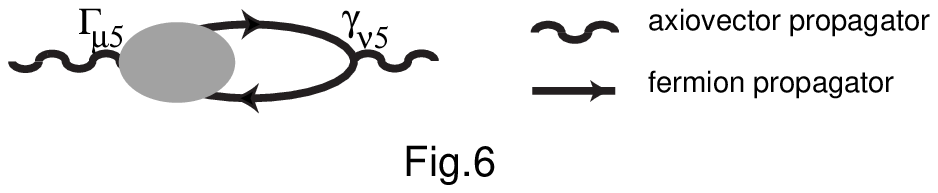}

This  example illustrates the fact that the transversality of the
polarization tensor used in the quantum field theoretic approach to mass
generation  is a consequence of a Ward identity.  This is true whether
vector masses  arise through fundamental fundamental BEH bosons  or
through fermion condensate.   The generation of gauge invariant masses is
therefore not contingent upon the ``tree approximation'' used to get the
propagators Eqs.(\ref{damass}) and (\ref{dnamass}). It is a consequence
of     the
$1/q^2$ singularity in the vacuum polarisation scalars Eqs.(\ref{pamass}),
(\ref{pnamass}) or (\ref{pdyn} ) which comes from  NG boson contribution.

\bigskip {\bf - The equation of motion approach} \cite {higgs1,higgs2}

Shortly after the above analysis was presented, Higgs wrote two papers.
In the first one \cite{higgs1} , he showed that the proof of the Goldstone
theorem
\cite{goldstone, gsw}, which states that, in relativistic quantum field
theory, spontaneous symmetry breaking of a continuous global symmetry
implies zero mass NG bosons, fails  in the case of gauge field theory. In
the second paper \cite{higgs2}, he  derived the BEH theory in terms of the
classical equations of motion, which he formulated for the abelian case.

From the action Eq.(\ref{local}), taking as in Eq.(\ref{vacuum}), the
expectation value of the BEH boson to be
$\langle\phi_1\rangle$, and expanding the NG field $\phi_2$ to first
order, one gets the classical equations of motion to that order
\begin{eqnarray}
\label{first} &&\partial^\mu \{\partial_\mu  \phi_2 -e\langle
\phi_1\rangle A_\mu\} =0\ ,\\
\label{second} &&\partial_\nu F^{\mu\nu}= e \langle
\phi_1\rangle\{\partial^\mu \phi_2 - e\langle
\phi_1\rangle A^\mu\}\ .
\end{eqnarray} 
 Defining
\begin{equation}
\label{hvector}
 B_\mu = A_\mu - {1\over e\langle
\phi_1\rangle}\partial_\mu 
\phi_2~~~\hbox{and}~~~G_{\mu\nu} =
\partial_\mu B_\nu -\partial_\nu B_\mu = F_{\mu\nu}\ ,
\end{equation} one gets
\begin{equation}
\label{vector}
\partial_\mu B^\mu =0 \ ,\quad\quad \partial_\nu G^{\mu\nu} + e^2 
\langle
\phi_1\rangle^2 B^\mu=0\ . 
\end{equation} Eq.(\ref{vector}) shows that $B_\mu$ is a massive vector
field with mass squared $ e^2 \langle\phi_1\rangle^2 $ in accordance with
Eq.(\ref{massa}).

In this formulation,    we see clearly how the Goldstone boson is absorbed
into a redefined massive vector field  which has no longer explicit gauge
invariance. The same phenomenon in the quantum field theory approach  is
related to the unobservability of the $1/q^2$ pole mentioned in the
discussion of Eq.(\ref{massna}); this will be made explicit in the next
section.

The equation of motion approach is classical in character but, as pointed
out by Higgs~\cite{higgs2}, the formulation of the
  BEH mechanism in the quantum field theory terms of
reference~\cite{eb} indicates its validity in the quantum regime. We now
show how the latter formulation signals the renormalizability
of the BEH theory.

\bigskip {\bf - The renormalization issue}

The massive vector propagator Eq.(\ref{dnamass}) differs from a
conventional free massive propagator in two respects. First the presence
of the unobservable longitudinal term reflects the arbitrariness of the
gauge  parameter
$\eta$. Second the NG pole at $q^2 =0$ in the transverse projector
$g_{\mu\nu}-q_\mu q_\nu / q^2$ is unconventional. Its significance is
made clear by expressing the propagator of the $A_\mu$ field in Eq.(\ref{dnamass}) as
(putting
$\eta$ to zero)
\begin{equation}
\label{dnamass2} D_{\mu\nu}^a
\equiv {g_{\mu\nu}-q_\mu q_\nu/q^2
\over q^2- \mu^{a2}}  ={g_{\mu\nu}-q_\mu q_\nu/\mu^{a2}
\over q^2- \mu^{a2}}  +{1\over \mu^{a2}} {q_\mu q_\nu
\over q^2}\ .
\end{equation} The first term in the right hand side of Eq.(\ref{dnamass2}) is the
conventional massive vector propagator.
It may be viewed as the (non-abelian
generalization of the) free propagator of the
$B_\mu$ field defined in Eq.(\ref{hvector}) while  the second term is  a
pure gauge propagator due to the NG boson ($[1/ e\langle
\phi_1\rangle]\partial_\mu 
\phi_2$  in Eq.(\ref{hvector})~) which converts the $A_\mu$ field into this massive vector
field $B_\mu$.

The propagator Eq.(\ref{dnamass}) which appeared in the field theoretic
approach contains thus, in the covariant gauges, the transverse projector
$g_{\mu\nu}- q_\mu q_\nu/q^2$ in the numerator of the massive gauge
 field $A_\mu^a$ propagator.   This is
in sharp contradistinction to the numerator
$g_{\mu\nu}-q_\mu q_\nu/\mu^{a2}$ characteristic of the conventional
massive vector field $B_\mu$ propagator. It is the transversality of  the self energy in
covariant gauges, which led in the ``tree approximation'' to the transverse projector in
Eq.(\ref{dnamass}). As already mentioned, the transversality is a consequence of a
Ward identity and  therefore does not depend on 
 the tree approximation. This fact is already suggested from the
dynamical example presented above but was proven in more general terms
in a subsequent publication\footnote{The proof given in reference
\cite{be2} was not complete because closed Yang-Mills loops, which
would have required the introduction of Fadeev-Popov ghosts were not
included.} \cite{be2}. The importance of this fact is that the transversality of the
self-energy in covariant gauges determines the power counting of irreducible
diagrams. It is then straightforward to verify that the BEH quantum field
theory formulation is renormalizable by power counting. 

On this basis we suggested that the BEH  theory constitutes indeed a consistent
renormalizable field theory \cite{be2}. To prove this statement, one must
verify that the theory is unitary, a fact which is not
apparent in the ``renormalizable'' covariant gauges because of the $1/q^2$ pole in the
projector, but would be  manifest in the ``unitary gauge'' defined in the free theory by the
$B_\mu$ propagator.  In the unitary gauge  however, renormalization from power counting 
is not  manifest. The equivalence,  at the free level, between the
 $A_\mu$ and  $B_\mu$ free propagators, which is only true in a gauge invariant theory
where their difference is the unobservable NG propagator appearing in Eq.(\ref{dnamass2}),
is the clue of the consistency of the BEH theory. A full proof that the theory is
renormalizable and unitary was achieved by 't Hooft and Veltman~\cite{renorm}. 

\bigskip\bigskip
 {\bf \large IV. Consequences}
\medskip

The most dramatic application of the BEH mechanism is the electroweak
theory, amply confirmed  by experiment. Considerable work has been done,
using the BEH mechanism,  to  formulate  Grand Unified theories of non
gravitational interactions. We shall summarize here these well known
ideas and then evoke the construction of regular monopoles and flux lines
using BEH bosons, because they raise potentially important conceptual
issues. We shall also mention briefly the attempts to include gravity in
the unification quest, in the so called M-theory approach, and focuses in
this context on   an  interesting geometrical interpretation of the BEH 
mechanism. 
\bigskip

{\bf - The electroweak theory} \cite{gws}

 In the electroweak theory, the gauge group is taken to be
$SU(2)\times U(1)$ with corresponding generators and coupling constants
$gA_\mu^aT^a$ and $g^\prime B_\mu Y^\prime $. The $SU(2)$ acts on
left-handed fermions only. The electromagnetic charge operator is
$Q=T^3 + Y^\prime$ and the electric charge $e$ is usually expressed in
terms of the mixing angle $\theta$ as
$g=e/\sin\theta  , g^\prime=e/\cos\theta $. The BEH bosons
$( \phi^+,\phi^0)$ are in a doublet of $SU(2)$ and their
$U(1)$ charge is $Y^\prime = 1/2$.  Breaking occurs in such a way that $Q$
generates an  unbroken subgroup, coupled to which is  the massless photon
field. Thus the vacuum is characterized by $
\langle\phi\rangle={1/\sqrt2}\  ( 0, v) $.

Using Eqs.(\ref{massa}) and (\ref{massna}) we get the mass matrix
\begin{center}
$\vert\mu^2\vert$=$\displaystyle{{v^2\over 4}} ~\begin{array}{|cccc|}
g^2&0&0&0\\ 0&g^2&0&0\\ 0&0&g^{\prime 2}& -g g^\prime\\ 0&0&-g
g^\prime & g^2
\end{array}$
\end{center} whose diagonalization yields the eigenvalues
\begin{equation}
M^2_{W^+}={v^2\over 4}g^2~,~ M^2_{W^-}={v^2\over
4}g^2~,~M^2_Z={v^2\over 4}~(g^{\prime 2}+ g^2)~,~M^2_A= 0 \ . 
\end{equation}
This permits to relate $v$ to  the the Fermi coupling $G$ as
$v^2= ({\sqrt 2} G)^{-1}$. 

Although the electroweak theory has been amply verified by experiment,
the existence of the BEH boson has, as yet, not been confirmed. It should
be noted that the physics of the BEH boson is more sensitive to dynamical
assumptions than the massive vectors $W^\pm$ and 
$Z$, be it a genuine elementary field or a manifestation of a composite due
to a more elaborate mechanism. Hence observation of its mass and width
is of particular interest for further understanding of the mechanism at
work. 

\bigskip 

{\bf - Grand unification schemes}

The discovery that confinement could be explained by the strong coupling
limit of quantum chromodynamics based on the ``color'' gauge group
$SU(3)$ led to tentative Grand Unification schemes where electroweak 
and strong interaction could be unified in a simple gauge group ${\cal G}$
containing $
 SU(2)\times U(1)\times SU(3)$  \cite{gqw}. Breaking occurs through
vacuum expectation values of BEH fields and unification can be realized at
high energies because while the renormalization group makes the small
gauge coupling of
$U(1)$  increase logarithmically with the energy scale, the converse is
true for the asymptotically free non abelian gauge groups. 
\bigskip

{\bf - Monopoles, flux tubes and electromagnetic duality}

In electromagnetism, monopoles can be included at the expense of
introducing a Dirac string \cite {dirac}. The latter creates a singular
potential along the string  terminating at the monopole.   For instance to
describe a point-like monopole located at
$\vec r =0 $, one can take the line-singular potential
\begin{equation}
\vec A = {g\over 4\pi} ( 1-\cos \theta) 
\vec \nabla \phi \ ,
\end{equation} This potential has a singularity along the negative $z$-axis
$(\theta =\pi) $ where the string has been put (see Fig.7). The
unobservability of the string implies that its fictitious flux be quantized
according to the Dirac condition
\begin{equation}
\label{dirac} eg =2\pi n\ \quad n\in Z \ .
\end{equation} 
\hskip 4cm \epsfbox{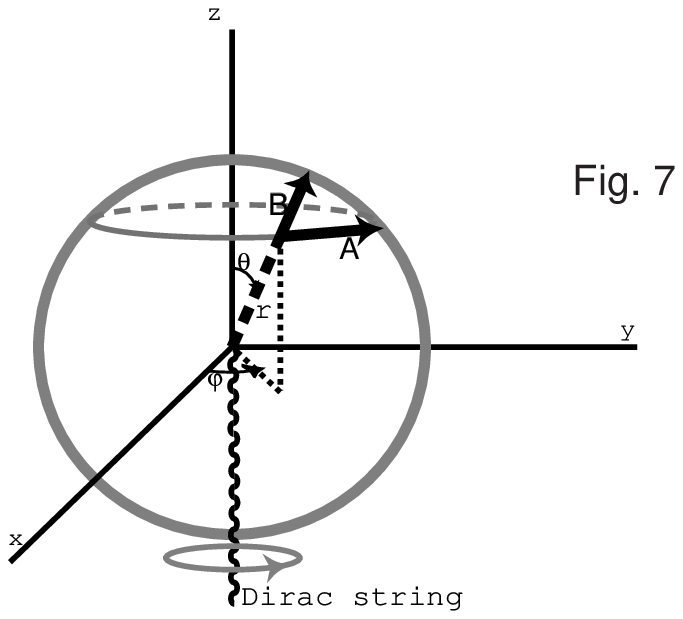}

In contradistinction to the string in the $U(1)$ theory, the
Dirac string in non abelian gauge groups can be removed by a gauge
singularity for well chosen quantized magnetic charges, reducing the line
singularity to a point like singularity. 

\hskip 1cm\epsfbox{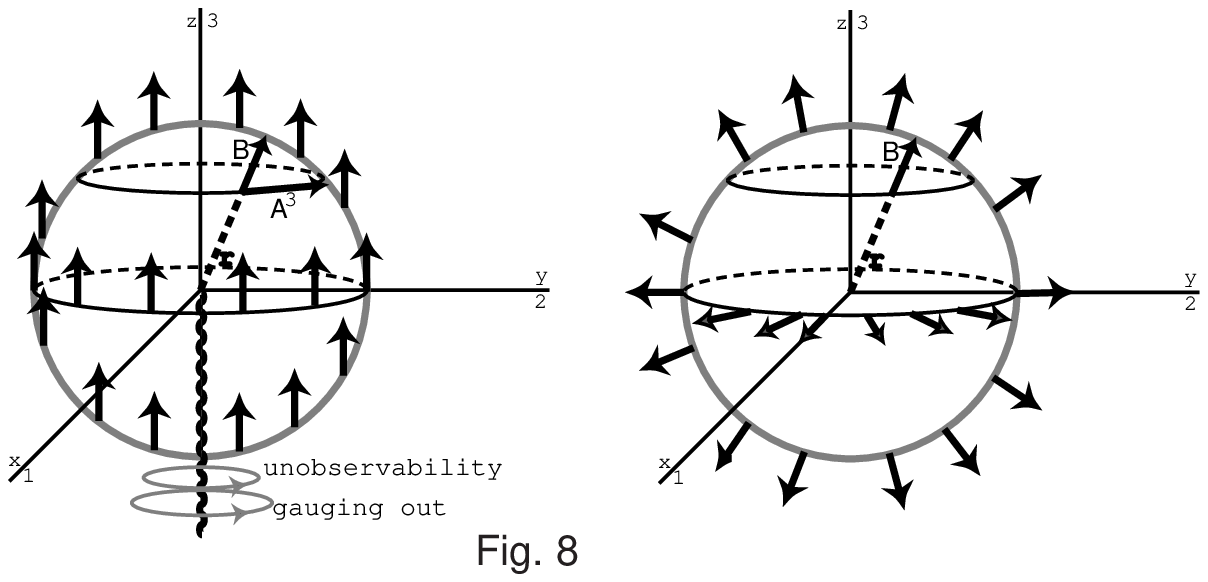}

An example is the SO(3) monopole, represented in Fig.8,
arising from the potential
\begin{equation}
\label{regular} A^{a\, i}={g\over 4\pi} \epsilon^{i a b} { r^b\over r^2}\quad ,
\quad eg =4\pi \ .
\end{equation} Breaking the symmetry to $U(1)$ by a BEH field belonging
to the adjoint group $SO(3)$ one can remove the point singularity to get
the topologically stable 't Hooft-Polyakov regular monopole \cite {hp}. 

This procedure can be extended to Lie  groups $\cal G$ of higher rank \cite
{all}. For a general Lie group $\cal G$, the possibility of gauging out the
Dirac string depends on the global properties of $\cal G$. Namely, the
mapping of a small circle surrounding the Dirac string onto $\cal G$ must
be a curve continuously deformable to zero. Closed curves in $\cal G$ are
characterized by $Z$  where
$Z$ is the subgroup of the center of the universal covering $
\tilde{\cal G}$ of 
${\cal G}$ such that
${\cal G}=\tilde{\cal G}/Z$. Gauging out only occurs for the  curve
corresponding to the unit element of $Z$. This is the origin for the
unconventional factor of $2$ ($4\pi =2.2\pi$) in Eq.(\ref{regular}) as
$SO(3)=SU(2)/Z_2$.

The construction of regular monopoles has  interesting conceptual
implications. 

The mixing between space and isospace indices in Eq.(\ref{regular}) means
that the regular monopole is invariant under the diagonal subgroup of
$SO(3)_{space}\times SO(3)_{isospace}$. This implies that a bound state
of a scalar of isospin $1/2$ with the monopole is a space-time fermion. In
this way, fermions can be made out of bosons \cite{fermions}.

One can define  regular monopoles in a limit in which the BEH-potential
vanishes. These are the BPS monopoles. They admit a supersymmetric
extensions in which there are indications that electromagnetic duality
can  be realized at a fundamental level, namely that the interchange of
electric and magnetic charge could be realized by equivalent but distinct
actions.

The BEH-mechanism, when $\cal G$ symmetry is completely broken,  is a
relativistic analog of superconductivity.  The latter may be viewed as a
condensation of electric charges. Magnetic flux is then channeled into 
quantized  flux tubes.  In confinement, it is the electric flux which is
channeled into quantized tubes. Therefore electric-magnetic duality
suggests that, at some fundamental level,  confinement is  a condensation
of magnetic monopoles and constitutes the magnetic dual of the BEH
mechanism
\cite{dual}.
\bigskip

{\bf - A geometrical interpretation of the BEH mechanism}

The BEH mechanism operates within the context of gauge theories. Despite
the fact that grand unification schemes reach scales comparable to the
Planck scale, there was, a priori, no indication that Yang-Mills fields
offer any insight into    quantum gravity. The only approach to quantum
gravity which had some success, in particular in the context of a quantum
interpretation of the black holes entropies,  are the superstring theory
approaches and the possible merging of the five perturbative approaches
(Type IIA, IIB, Type I and the two heterotic strings) into an elusive
M-theory whose classical limit would be 11-dimensional supergravity.  Of
particular interest  in that context is the discovery  of Dp-branes along 
which the ends of  open strings  can  move \cite{pol}. This led, for the
first time, to an interpretation of the area entropy of some black holes in
terms of a counting of quantum states. Here we shall explain how Dp-branes
yield a geometrical interpretation of the BEH mechanism.

When $N$  BPS Dp-branes coincide, they admit  massless excitations  from
the 
$N^2$ zero length oriented strings with both end attached on the $N$
coincident branes. There are $N^2$ massless vectors and additional $N^2$
massless scalars for each dimension transverse to the branes. The open
string sector has local $U(N)$ invariance. At rest,  BPS Dp-branes  can
separate from each other in the transverse dimensions   at no cost of
energy. Clearly this can break the symmetry group from $U(N)$ up to
$U(1)^N$ when all the branes are at distinct location in the transverse
space, because strings joining two different branes have finite length and
hence now describe finite mass excitations.  The only remaining massless
excitations are then due to the zero length strings with both ends  on the
same brane.  

\hskip 4cm\epsfbox{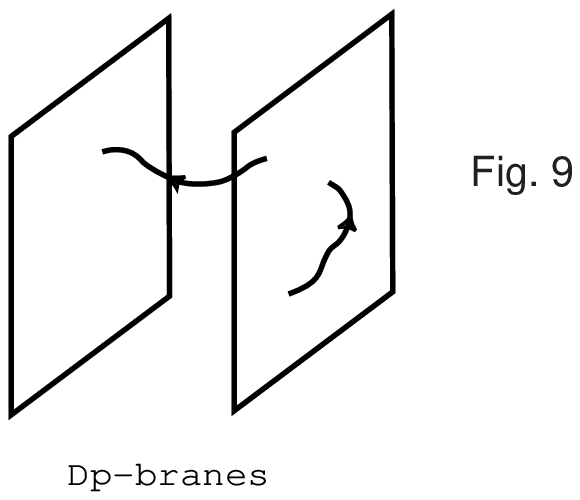}

This symmetry breaking mechanism can be understood as a BEH mechanism from the  action
describing   low energy excitations of
$N$ Dp-branes. This  action  is the  reduction to
$p+1$ dimensions of 10-dimensional supersymmetric Yang-Mills with
$U(N)$ gauge  fields \cite {redu,reduc}. 

The Lagrangian is
\begin{equation}
\label{reduction} {\cal L}=- {1\over4}Tr {\bf F}_{\mu\nu} {\bf
F}^{\mu\nu}+  Tr
\left(~  {1\over2} D_\mu{\bf A}^iD^\mu{\bf A}^i - {1\over4} [{\bf A}^i\, ,
{\bf A}^j]^2  ~\right)+
\hbox{fermions} \ ,
\end{equation} where $\mu$ labels the $p+1$ brane coordinates and $i$
the directions transverse to the branes. ${\bf F}_{\mu\nu}=F_{\mu\nu}^a
{\bf T}^a $, ${\bf A}^i=A^{a\, i}~ {\bf T}^a$ where ${\bf T}^a$ is a generator
of $U(N)$ in a defining representation.

The states of zero energy are given classically, and hence in general
because of supersymmetry, by all commuting
${\bf A}^i =\{x^i_{mn}\}$  matrices, that is, up to an equivalence, by all
diagonal matrices $\{x^i_{mn}\}=\{x^i_{m}\delta_{mn}\}$. Label the $N^2$
matrix elements of ${\bf A}_\mu$ by $A_{\mu\, mn}$. The 
$(N^2-N)$ gauge  fields given by the non diagonal elements $m\neq n$
acquire a mass
\begin{equation}
\label{nd} m^2_{mn}\propto (\vec x_m -\vec x_n)^2 \ ,
\end{equation} if $\vec x_m\neq\vec x_n $, as   is easily checked by computing  the
quadratic  terms in $A_{\mu\, mn}$ appearing in the covariant derivatives $Tr D_\mu{\bf
A}^iD^\mu{\bf A}^i$. 

This symmetry breaking is
 induced by  the  expectation values 
$\{x^i_{m}\}$. The gauge invariance is ensured, as usual, by unobservable
($N^2-N$) NG bosons. To identify the latter we consider the scalar
potential in Eq.(\ref{reduction}), namely
\begin{equation}
\label{scalar}
V= Tr {1\over4} [{\bf A}^i\, ,
{\bf A}^j] [{\bf A}^i\, ,
{\bf A}^j] = {1\over4}\sum_{i,j;m,n} \langle m\vert [{\bf A}^i\, ,
{\bf A}^j] \vert n \rangle \langle n\vert [{\bf A}^i\, ,
{\bf A}^j]\vert m\rangle\, .
\end{equation}
We write
\begin{equation}
\langle m\vert {\bf A}^j \vert n \rangle = x^j_{m}\delta_{mn} + y^j_{mn}\, .
\end{equation}
Here the   diagonal elements
$\{x^j_{m}\}$ are the BEH expectation values and the  $y^j_{mn}(= - [y^j_{nm}]^*)$
define
$d(N^2-N)$ hermitian scalar fields  $(y^i_{mn})^a ~ (a= 1,2)$ where $y^j_{mn} = (y^j_{mn})^1
+ i (y^j_{mn})^2 ~, ~ m>n$~,  and
$d$ is the number of transverse space dimensions. The mass matrix for the fields
$(y^i_{mn})^a $ is
\begin{equation}
\label{matrix}
{\partial ^2 V\over \partial (y^k_{mn})^a \partial (y^l_{mn})^b }= \delta^{ab}[(\vec x_m -\vec
x_n)^2
\delta^{kl}- (x^k_{m}-x^k_{n})(x^l_{m}-x^l_{n})]\, ,
\end{equation}
and has for each  pair $m,n ~(m<n)$, two zero eigenvalues corresponding to
the eigenvectors  
$ (y^l_{mn})^a\propto (x^l_{m}-x^l_{n})$. These are the required $(N^2-N)$ NG
bosons, as can be checked directly from the coupling of ${\bf A}^i$ to  ${\bf
A}_\mu$ in the Lagrangian Eq.(\ref{reduction})~.

As mentioned above, the breaking of $U(N)$ up to
$U(1)^N$  may be viewed in the string picture
 as due to  the stretched  strings joining branes separated in the
 dimensions transverse to the branes. One identifies the $\{x^i_{m}\}$ as 
 coordinates transverse to the brane
$m$. The mass of  the vector meson $A_{\mu\, mn}$ is then the mass shift
due to the stretching  of the otherwise massless  open string
vector excitations.   The unobservable  NG bosons $\vec
y_{mn}\parallel (\vec x_m -\vec x_n)
$ are  the field theoretic expression of the unobservable longitudinal
modes of the strings joining the branes $m$ and $n$.   In this  way
Dp-branes provide a geometrical interpretation of the BEH mechanism.

It may be worth  mentioning the  interesting situation which occurs when
$p=0$ \cite {reduc, bfss}. The Lagrangian Eq.(\ref{reduction}) then
describes
 a pure quantum mechanical system where the $\{x^i_{mn}\}$ are the
dynamical variable. The time component 
${\bf A}_t$ which enters the covariant derivative $D_t {\bf A}^i$ can be
put equal to zero, leaving a constraint which amounts to restrict the
quantum states to singlets of $SU(N)$. The $\{x^i_{m}\}$ which define in
string theory D$0$-brane coordinates (viewed as partons in the infinite
momentum frame in reference~\cite{bfss}) are  the analog, for $p=0$, of 
the BEH expectation values in the $p\neq0$ case, although  they label now
classical collective position variables of the quantum mechanical system
and not vacuum expectation values. The nondiagonal quantum degrees of
freedom 
$\vec y_{mn}~\bot~ (\vec x_m -\vec x_n)$  have  a positive potential 
energy proportional to the distance squared between the D$0$-branes $m$
and
$n$.  Hence they get locked in their ground state when the D$0$-branes are
 largely separated  from each other.  In this way, the
D0-brane
${\bf A}^i =\{x^i_{mn}\}$  matrices  commute at large distance
scale and define geometrical degrees of freedom.
However these  matrices  do not commute at short distances where the
potential energies of the
$y^i_{mn}$ go to zero. This suggests that the space-time geometry
exhibits non commutativity at small distances, a feature which may well
turn out to be an essential element of  quantum gravity.

\bigskip\bigskip

{\bf \large V. Remarks}
\medskip

Physics, as we know it, is an attempt to interpret  the apparent diversity
of natural phenomena in terms of general laws. By essence then, it incites
one towards a quest for unifying  diverse physical laws.

Originally the BEH mechanism was conceived to unify the theoretical
description of long range and short range forces. The success of the
electroweak theory made the mechanism a candidate for further
unification. Grand unification schemes,  where the scale of unification is
pushed close to the scale of quantum gravity effects, raised the
possibility that unification might also have to include gravity. This trend
towards the quest for unification received a further impulse from the
developments of string theory  and from its connection  with
eleven-dimensional supergravity. The latter was then viewed as a
classical limit of a hypothetical M-theory into which all perturbative
string theories would merge. In that context, the geometrization of the
BEH mechanism is suggestive of the existence of an underlying non
commutative geometry.

\end{document}